\documentclass[12pt,a4paper]{article}
\usepackage[cp1251]{inputenc}
\inputencoding{cp1251}
\usepackage[english]{babel}

\def\bb{\begin{equation}}
\def\ee{\end{equation}}

\newtheorem{theorem}{Theorem}
\newtheorem{lemma}{Lemma}

\title{\bf{Scattering problem for the local parametric resonance equation}
\footnote{This work was supported by RFBR grants 06-01-00124-a and 06-01-92052-KE-A and DFG project GZ: TA 289/4-1}}

\author{Oleg Kiselev\footnote{Institute of Mathematics USC RAN; ok@ufanet.ru},\and Yulia Bagderina${}^\dag$,\and Sergei Glebov\footnote{Ufa State Petroleum Technical University; sg@anrb.ru}}

\date{\null}
\begin{document}
\maketitle

\begin{abstract}
In this paper we present the solution of local parametric resonance equation in terms of parabolic cylinder functions and solve the scattering problem.  
\end{abstract}

\section{Introduction}\label{sectionIntroduction}
\par

In this work we present a solution of scattering problem for an equation
\bb
i {dw\over dt}+t w+ a\overline{w}=0.
\label{localResonanceEq}
\ee
here $\overline{w}$ is a complex conjugate function of $w$,  $a$ is a real constant.
\par
If we  use the constant coefficient $T$ instead of $t$ in the second term of equation (\ref{localResonanceEq}) then such equation describes a well-known parametric resonance phenomenon \cite{BMitr}. In our case equation (\ref{localResonanceEq}) determines an amplitude of oscillating solution for a local parametric resonance. The most interesting problem for  applications is a scattering problem for (\ref{localResonanceEq}). The statement of the problem is as follows: we will determine the asymptotic behavior of solution for (\ref{localResonanceEq})  when $t\to\infty$ with the given asymptotic behavior when  $t\to-\infty$.
\par
Firstly the local nonparametric resonance was studied in \cite{Kev}, \cite{Abl}.  It was shown that the passage through the local resonance is described by the Fresnel integral. The 	analogous formula for the parametric resonance was unknown up to now.  Here we present it.
\par
Well-known handbooks  contain the explicit expression neither for a solution of system 
 for real and imaginary parts of solution  $w=x+iy$
\begin{eqnarray}
{dx\over dt}=(a-t)y,\nonumber\\
{dy\over dt}=(a+t)x.
\label{realaormEq}
\end{eqnarray}
nor for the second order differential equation that is equivalent to this system
\bb
x''+(t^2-a^2)x+{x'\over t-a}=0.
\label{secondOrderEq}
\ee
\par
In this paper we present the formula for the solution of system (\ref{realaormEq}) and close this gap. 

\section{Result}
\par
\begin{theorem}\label{theorAboutSolution}
The solution of equation  (\ref{localResonanceEq}) is represented by  parabolic cylinder functions
\begin{eqnarray}
w(t,a)={1\over2}e^{-i\frac{3\pi}{4}-\frac{3\pi a^2}{8}}
\, a\, \overline{U} D_{z}(e^{i\pi\over4}\sqrt{2} t) + \nonumber\\
\bigg(e^{\frac{\pi a^2}{8}}U+ 2e^{i\frac{3\pi}{4}+\frac{\pi a^2}{8}}
{\sqrt{2\pi}\over a\Gamma(-i{a^2\over2})}\overline{U}\bigg)D_{-z-1}(e^{{i5\pi\over4}}\sqrt{2} t).
\label{genSolution}
\end{eqnarray}
Here  $z=i\displaystyle{a^2\over2}-1$, the function  $D_n(\zeta)$  is the parabolic cylinder function
 \cite{WhittakerWatson}, the parameter  $U$  is an arbitrary complex constant. 
\end{theorem}
\par{\bf Corollary.} Let the solution of (\ref{localResonanceEq})  be 
\bb
w=V e^{i\big({t^2\over2}-{a^2\over2}\ln(-t)\big)}(1+O(t^{-1})),\quad \forall V\in{\mathbf C},\quad t\to-\infty,
\label{leftAs}
\ee
then  this solution has the following form as  $t\to\infty$
\bb
w=\bigg(e^{{a^2\pi\over2}} V+{e^{i\frac{\pi-2a^2\ln(2)}{4}+\frac{\pi a^2}{4}}  
a\sqrt{\pi}\over \Gamma(1-i{a^2\over2})} \overline{V}\bigg)e^{i\big({t^2\over2}-{a^2\over2}\ln(t)\big)}(1+O(t^{-1})).
\label{rightAs}
\ee
Formulas  (\ref{leftAs}) and  (\ref{rightAs}) give the connection formulas for solutions of equation  (\ref{localResonanceEq}).

\section{Proof of Theorem  \ref{theorAboutSolution}}
\par
The proof of Theorem  \ref{theorAboutSolution} consists of  three steps. The first step shows that the equation of parabolic cylinder is a differential consequence of equation (\ref{localResonanceEq}). It yields that the solution of  (\ref{localResonanceEq})  can be represented as a linear combination of  known solutions  of parabolic cylinder equation. The second step contains the construction of WKB-asymptotic representation for the solution of 
(\ref{localResonanceEq}) when $t\to-\infty$. The final  step allows us to obtain formulas that represent the solution of (\ref{localResonanceEq})  by linear independent solutions of parabolic cylinder equation. 

\subsection{Reduction to the parabolic cylinder equation}
\begin{lemma}\label{lemmaParabolicCylinder}
The solution of equation (\ref{localResonanceEq}) is a  combination of linearly independent solutions of parabolic cylinder equation with special values of constants  $c_1$ and  $c_2$:
\bb
w(t,a)=c_1D_n(e^{i{\pi\over4}}\sqrt{2}t)+ c_2  D_{-n-1}(e^{i{3\pi\over4}}\sqrt{2}t),
\label{solution}
\ee
where  $n=-1+ia^2/2$.
\end{lemma}
To prove Lemma \ref{lemmaParabolicCylinder} we differentiate equation (\ref{localResonanceEq}) with respect to $t$. It yields the differential equation of the second order. We change the derivative $\overline{w}'$ according to the complex conjugated equation of (\ref{localResonanceEq}) and $\overline{w}$ according to equation (\ref{localResonanceEq}).
 It yields the following equation
\bb
w''-(i+a^2-t^2)w=0.
\label{parabolicCylinderEq}
\ee
Denote  $W(\zeta)=w(t,a)$, where $\zeta=e^{i{\pi\over4}}\sqrt{2}t$, $n=-1+ia^2/2$. This substitution leads to  the standard form of parabolic cylinder equation for function $W(\zeta)$
$$
{d^2 W\over d\zeta^2}+\big(n+{1\over2}-{1\over4}\zeta^2\big)W=0.
$$ 
A general solution of this equation can be represented as a  combination of linear independed solutions
$$
W=c_1 D_n(\zeta)+c_2 D_{-n-1}(i\zeta).
$$
The solution of equation (\ref{localResonanceEq}) satisfies equation (\ref{parabolicCylinderEq}) as well and can be represented in form (\ref{solution}).  Lemma \ref{lemmaParabolicCylinder} proved.

\subsection{WKB-asymptotic representation}
\begin{lemma}\label{lemmaWKB}
When $t\to-\infty$ the solution of equation  (\ref{localResonanceEq}) has the form:
\bb
w=e^{i\big({1\over2}t^2-{1\over2}a^2\ln(-t)\big)}U+{1\over t} e^{-i\big({1\over2}t^2-{1\over2}a^2\ln(-t)\big)}a\overline{U}+O(t^{-2}).
\label{WKBAsymptotic}
\ee
\end{lemma}
The proof of Lemma \ref{lemmaWKB} can be obtained by the standard way. We construct the formal asymptotic solution of the form
$$
w=e^{i\big(\Omega(t)+\omega\ln(-t)\big)}\sum_{k=0}^{\infty}t^{-2k}\phi_k +e^{-i\big(\Omega(t)+\omega\ln(-t)\big)}\sum_{k=1}^{\infty}t^{-2k+1}\psi_k.
$$
Substitute this formula into equation (\ref{localResonanceEq}) and gather the terms of the same order with respect to $t$. It yields the recurrent sequence of equations for $\Omega(t), \omega$ and coefficients of asymptotic solution.
\par
Relations of order  $t$  give
$$
(-\Omega'+ t)\phi_0e^{i\big(\Omega(t)+\omega\ln(-t)\big)}=0,
$$
and hence  $\Omega=t^2/2$.
\par
Relations of order $t^0$  have the form
$$
(2\psi_1+a\overline{\phi_0}) e^{-i\big(\Omega(t)+\omega\ln(-t)\big)}=0,
$$
and  hence $\psi_1=-a\overline{\phi_0}/2$.
\par
Relations of order $t^{-1}$ have the form
$$
(-\omega- {1\over2}|a|^2)\phi_0 e^{i\big(\Omega(t)+\omega\ln(-t)\big)}=0,
$$
and give $\omega=-|a|^2/2$.
\par
Denote  $\phi_0=U$, where $U\in {\mathbf C}$ and obtain the leading-order terms of the asymptotic representation  of (\ref{WKBAsymptotic}). The higher order terms are evaluated in the same manner. The justification of these asymptotic formulas is realized by the standard way, see \cite{Wasov}. Lemma \ref{lemmaWKB} proved.

\subsection{Matching of asymptotic representations}
\par
In this section we match asymptotic formula (\ref{WKBAsymptotic}) and formula of the solution (\ref{solution}). To obtain this result we use well-known formulas for asymptotic representation of parabolic cylinder functions \cite{WhittakerWatson}. Here we present these formulas for reference 
\begin{eqnarray}
D_n(\zeta)=& e^{-{1\over4}\zeta^2} \zeta^n\big(1-O(\zeta^{-2})\big), &\quad |\arg(\zeta)|<{3\over4}\pi;\nonumber\\
D_n(\zeta)=& e^{-{1\over4}\zeta^2} \zeta^n\big(1-O(\zeta^{-2})\big)- & \quad \nonumber\\
&{\sqrt{2\pi}\over \Gamma(-n)}e^{i n\pi}\zeta^{-n-1}\big(1+O(\zeta^{-2})\big), &\quad {1\over4}\pi<\arg(\zeta)<{5\over4}\pi;\nonumber\\
D_n(\zeta)= &e^{-{1\over4}\zeta^2} \zeta^n\big(1-O(\zeta^{-2})\big)-& \quad \label{asForPCF} \\
&{\sqrt{2\pi}\over \Gamma(-n)}e^{-i n\pi}\zeta^{-n-1}\big(1+O(\zeta^{-2})\big), &\quad -{5\over4}\pi<\arg(\zeta)<-{1\over4}\pi.\nonumber
\end{eqnarray}
Substitute asymptotic representation (\ref{WKBAsymptotic}) into the left-hand side of (\ref{solution}) and asymptotic representation (\ref{asForPCF})  of parabolic cylinder functions into the right-hand side of (\ref{solution}). Gathering  the terms of the same order with respect to  $t$ gives the statement of Theorem \ref{theorAboutSolution}. Theorem \ref{theorAboutSolution} proved.
\par
The corollary of the Theorem \ref{theorAboutSolution} is obtained from the general solution (\ref{genSolution}) of equation (\ref{parabolicCylinderEq})  and known asymptotic representation (\ref{asForPCF}) for parabolic cylinder functions.
\par
{\bf  Acknowledgments.}  We wish like to thank B.I. Suleimanov for helpful comments.

\end{document}